# Flat bands and temperature-driven phase transition in quasi-one-dimensional zigzag chains

Jisong Gao,[1,2,#] Haijun Cao,[1,2,#] Xuegao Hu,[1,2,#] Hui Zhou[1,2], Zhihao Cai,[1,2] Qiaoxiao Zhao,[1,2] Dong Li,[1,2] Zhicheng Gao,[1,2] Shin-ichiro Ideta,[3] Kenya Shimada,[3] Peng Cheng,[1,2] Lan Chen,[1,2,4] Kehui Wu,[1,2,4,5*] Sheng Meng,[1,2,4,5*] Baojie Feng,[1,2,4,5*]

[1]*Institute of Physics, Chinese Academy of Sciences, Beijing, 100190, China*

[2]*School of Physical Sciences, University of Chinese Academy of Sciences, Beijing, 100049, China*

[3]*Hiroshima Synchrotron Radiation Center, Hiroshima University, Higashi-Hiroshima 739-0046, Japan*

[4]*Songshan Lake Materials Laboratory, Dongguan, Guangdong, 523808, China*

[5]*Interdisciplinary Institute of Light-Element Quantum Materials and Research Center for Light-Element Advanced Materials, Peking University, Beijing, 100871, China*

[#]These authors contributed equally to this work.

*Corresponding author. E-mail: khwu@iphy.ac.cn; smeng@iphy.ac.cn; bjfeng@iphy.ac.cn.

**Abstract**

Flat-band materials have garnered extensive attention due to their captivating properties associated with strong correlation effects. While flat bands have been discovered in several types of 2D materials, their existence in 1D systems remains elusive. Here, we propose a 1D frustrated lattice, specifically the 1D zigzag lattice, as a platform for hosting flat bands. This lattice can be experimentally realized by growing CuTe chains on Cu(111). The presence of flat bands was confirmed by tight-binding model analysis, first-principles calculations, and angle-resolved photoemission spectroscopy measurements. In addition, we discovered a temperature-driven phase transition at approximately 250 K. Detailed analyses demonstrate that the system has a Tomonaga-Luttinger liquid behavior, accompanied by spin-charge separation effects. Our work unveils new prospects for investigating strongly correlated electron behaviors and topological properties in the 1D limit.

Flat-band materials are characterized by a dispersionless energy-momentum relationship that leads to a highly degenerate density of states and quenched electron kinetic energies. These unique electronic structures give rise to a range of physical phenomena driven by electron correlations, including high-temperature superconductivity [1],[2], fractional quantum Hall effects [3], and Wigner crystallization [4]. To date, limited types of two-dimensional (2D) systems have been confirmed to possess flat bands. One is the kagome lattice, including $Fe_3Sn_2$, CoSn, FeSn, and $Nb_3Cl_8$, wherein the flat bands arise from the destructive interference of electron wavefunctions [5]-[10]. Another system is the moiré system, in which electrons are spatially confined by a long-range moiré potential [11]-[13]. In addition, flat bands also exist in Mott insulators, such as 1T-$TaS_2$ [14], which only emerge at low temperatures.

Due to the trend toward device miniaturization, 1D materials, such as carbon nanotubes [15] and graphene nanoribbons [16], have garnered considerable interest in recent decades. However, experimental evidence of flat bands in 1D materials is still scarce, despite the recent prediction of electronic 1D flat bands in several moiré systems [17],[18]. Regarding the real 1D materials, several toy models, including the diamond, cross, stub, and sawtooth lattices, have been extensively explored theoretically [19]-[27], with material prediction and realization remaining difficult. Additionally, even without flat bands, the reduced dimensionality in 1D materials can strengthen the Coulomb interaction of electrons [28]. Since strong correlation effects can lead to a variety of physical phenomena, such as Tomonaga-Luttinger liquid (TLL) behavior, charge and spin density waves, Peierls instability, fractional quantum Hall effect, and spin-charge separation [29]-[31], the addition of a flat band in 1D materials can further enhance the correlation effects, potentially giving rise to even more exotic properties. As a result, 1D materials with flat bands provide a unique platform to investigate the interplay between electron-correlation effects, lattice symmetry, and quantum confinements.

Here, we present the realization of 1D flat bands in CuTe chains, which can be simplified into the zigzag lattice model. Our angle-resolved photoemission spectroscopy (ARPES) measurements provide direct evidence of the presence of flat bands, which is backed up by our scanning tunneling microscopy/spectroscopy (STM/S) measurements, first-principles calculations, and tight-binding model analysis. In addition, we observed an unexpected phase transition at approximately 250 K, which might originate from the spin-charge separation effects. These results together indicate the rich correlation physics in the CuTe chains.

A well-known 1D lattice that are expected to host flat bands is the sawtooth lattice, as schematically shown in Fig. 1(a). Due to the destructive interference of electron wavefunctions, a completely flat band exists across the entire Brillouin zone (BZ), accompanied by an electron-like bands [see Fig. 1(c)]. However, in realistic materials, the hopping between neighboring sites in the top row, such as sites 2 and 3, cannot be ignored due to similar chemical environments of the top and bottom atoms. This results in the formation of a 1D zigzag lattice, as shown in Fig. 1(b). Two hopping terms, $t$ and $t'$, correspond to the hopping of an electron from one site to its nearest neighbor along the two different bond directions. The zigzag lattice can exhibit an ideal flat band over half of the BZ with suitable hopping amplitudes, as shown in Fig. 1(d) (see also Fig. S1). As the BZ boundary is approached, the flat band disperses to higher binding energies and touches the electron-like band at the boundary of the BZ, giving rise to a 1D Dirac cone [32],[33]. The band degeneracy is protected by the glide mirror symmetry of the zigzag lattice and cannot be eliminated by varying hopping amplitudes.

The 1D zigzag lattice can be realized in CuTe chains which can be grown on Cu(111). The preparation method and atomic structure of CuTe/Cu(111) have been reported recently [34],[35], but its topological properties have not been studied. Schematic drawings of the atomic structure of CuTe/Cu(111) are displayed in Figs. 1(e) and 1(f). As the Cu and Te atoms are not coplanar, the hybridization between the in-plane orbitals of Cu and Te is small. Therefore, the Cu atoms form a 1D zigzag lattice with little perturbations from the Te atoms. Due to the 1D nature of the CuTe chains and the three-fold symmetry of the Cu(111) substrate, three equivalent domains coexist, forming a $3\times\sqrt{3}$ superstructure with respect to the 1×1 lattice of Cu(111) in the LEED patterns, as shown in Fig. S2.

Having prepared high-quality CuTe chains, we conducted ARPES measurements to investigate their electronic structures. On the Fermi surface, there is a discernible dot-like feature at each K point of Cu(111), as indicated by the black arrows in Fig. 2(a). At a binding energy of 1.5 eV, a strong spectral weight emerges at the boundary of the BZ of the $3 \times \sqrt{3}$ superstructure, forming a hexagonal shape due to the presence of three rotational domains, as highlighted by the red arrow in Fig. 2(b). Remarkably, the spectral weight is extremely straight at $E_B$=1.5 eV, and the straight feature disappears at other binding energies, suggesting the existence of a flat band at $E_B$=1.5 eV along CuTe chains.

To verify the flat band, we present ARPES intensity plots along Cuts 1 and 2 [see Fig. 3(a)] in Figs. 3(b) and 3(c), respectively. These two cuts correspond to the direction along the CuTe chains and are positioned at the boundaries of the first and second BZs of the $3 \times \sqrt{3}$ superstructure, respectively, as indicated by the black lines in Fig. 3(a). Neglecting the variation of the intensity of the spectral weight, we observe that the two cuts exhibit similar features, consistent with the periodicity in real space. The other two equivalent domains do not have such symmetries, which confirms that these bands originate from the parallel CuTe chains. Notably, a flat band at the binding energy of 1.5 eV is observed, which is absent in clean Cu(111) and independent of photon energy (Figs. S3 and S4). The flat band extends to approximately half of the BZ, which corresponds to the straight spectral weight on the constant energy contour (CEC) in Fig. 2(b). We further performed STM/S measurements and observed a prominent peak at approximately 1.5 eV below the Fermi level, which is in line with the flat band (see Fig. S5).

Approaching the boundary of the BZ, the otherwise flat band disperses to higher binding energies. This downward dispersion is reminiscent of the 1D Dirac cone at the BZ boundary of the zigzag lattice. However, the lower portion of the Dirac cone is mixed with the bulk bands of Cu(111) and is difficult to discern. Above the flat band, a hole-like band has been observed along the two cuts. The band top is located at ~0.4 eV below the Fermi level because the two cuts are away from the Γ point of the $3 \times \sqrt{3}$ superstructure.

To further confirm the electronic structures of CuTe chains, we performed first-principles calculations and unfolded the band structures to the first BZ of Cu(111) for better comparison with our ARPES results. Figure 3(d) shows the calculated band structures along Cut 1, revealing clear evidence of a flat and hole-like bands [see also Fig. S6]. Our calculation results agree well with our experimental results, despite the slight difference in chemical potential between our calculations and real materials. Notably, we found that the bands with high spectral weight in the calculated band structures [as indicated by the red arrows in Fig. 3(d)] arise from the finite thickness of our slab calculation and are sensitive to the number of layers of the Cu(111) substrate, as demonstrated in Fig. S7.

Figure 3(e) displays the density of states contributed by Te, Cu, and the substrate, respectively. The sharp peak at ~1.5 eV (indicated by the black arrow) is attributed to the Cu orbitals in CuTe chains and corresponds to the flat band observed in the calculated band structures. These results suggest that the flat band primarily originates from the Cu atoms in CuTe chains, rather than the substrate, which is consistent with the zigzag lattice model. To further investigate the orbital contributions to the flat band, we disentangle the contributions of different Cu orbitals, as shown in Fig. 3(f), which show that the flat band is primarily derived from the Cu $d_{x^2-y^2}$ orbitals. Figures 3(g) and 3(h) display the calculated band structures with only contributions from Cu atoms in CuTe chains and the $d_{x^2-y^2}$ orbitals of Cu, respectively, where the flat band is clearly observed. The $d_{x^2-y^2}$ orbitals of Cu are located in-plane and have little overlap with the Te orbitals that are not coplanar with the Cu atoms, which provides an explanation for the persistence of the flat band in the zigzag lattice formed by Cu atoms. On the other hand, the hole-like bands are primarily contributed by the Te and bulk Cu atoms which do not host flat bands. In addition, Te atoms are much closer to neighboring chains compared to Cu atoms, giving rise to a strong 2D dispersion.

To gain a deeper understanding of the origin of the flat band, we performed a tight-binding (TB) model analysis. Specifically, we focused solely on the Cu $d_{x^2-y^2}$ orbitals, given their dominant contribution to the flat band. The Cu(111) substrate interacts with the CuTe chain and modulates the hopping integrals and on-site energies. The Hamiltonian can be expressed as:

$$H = -t \sum_{i}(a_{i,1}^{\dagger} a_{i,2} + a_{i,1}^{\dagger} a_{i-1,2} + h.c.) - t' \sum_{i,s=1,2}(a_{i,s}^{\dagger} a_{i-1,s} + h.c.) + U \sum_{i,s=1,2} a_{i,s}^{\dagger} a_{i,s}$$

where $a_{i,s}^{\dagger}$ and $a_{i,s}$ denote the creation and annihilation operators for the Cu $d_{x^2-y^2}$ orbitals at the site $s \in \{1,2\}$ in the $i^{th}$ unit cell, $t$ and $t'$ the hopping integrals between neighboring atoms along different bond directions [as illustrated in Fig. 1(b)], and $U$ the on-site energy imposed by the Cu(111) substrate. By fitting this model to our experimental data, we obtained the following parameter values: $t = 0.4$ eV, $t' = 0.1$ eV, $U = -1.7$ eV. These parameters give rise to a flat band, as visualized in Fig. 1d, which is consistent with both our ARPES measurements and first-principles calculation results. More detailed discussion on the 2D TB model can be found in the Supplementary Materials.

Having confirmed existence of 1D flat bands in CuTe chains, we proceed to study the electronic states near the Fermi level, which can be renormalized by the strong electron correlation effects. We examine the band structure of CuTe/Cu(111) along the $\overline{KMK}$ direction of Cu(111). At 300 K, we observed a parabolic band, α, with its top close to the Fermi level, as shown in Figs. 4(a) and 4(b). This band is consistent with our DFT+U calculations, as shown in in Figs. 4(c) and S10. The estimated U value is 7 eV, which indicate the presence of strong electron correlation effects. Interestingly, an additional band, β, appears at 70 K, as shown in Figs. 4(b), which was not predicted by calculations. Detailed temperature-dependent ARPES measurements, presented in Figs. S11 and S12, reveal that the phase transition is reversible with both decreasing and increasing temperatures. Figure 4(d) displays the temperature evolution of the spectral weight of the two bands, alongside the bulk *sp* band of Cu(111) as a reference. We determine the critical temperature for the phase transition to be approximately 250 K, close to room temperature.

To elucidate the origin of this phase transition, we performed LEED and XPS measurements. No detectable changes in LEED patterns or shifts in XPS peaks were observed across the transition temperature, as shown in Fig. S13. Additionally, the β band exhibits the same periodicity as the α band, indicating the absence of structural phase transitions or long-range charge orders. We can also rule out the possibility of phonon-induced replica bands because the replica band typically exhibits a rigid energy shift relative to the original band [36,37]. The coexistence of the α and β bands at low temperatures is reminiscent of the spin-charge separation phenomenon, caused by strong electron correlation effects [38-40]. In this context, the α and β bands correspond to the spinon and holon branches, originating from spin and charge excitations, respectively. At 70 K, both branches are observable by ARPES. As the temperature increases, the spinon branch gains spectral weight due to the thermal excitation of spin states [41,42], resulting in the depletion of the holon branch. Above the critical temperature (~250 K), the holon branch completely disappears. The persistence of the spinon branch at higher temperatures demonstrates that the system continues to exhibit spin-charge separation. This scenario is consistent with our experimental results in Fig. 4(d), which show a simultaneous enhancement of the α band and the disappearance of the β band.

The reason for the abrupt phase transition, as opposed to a smooth transition, remains unclear and requires further experiments and calculations to determine the magnetic and spin-excited states.

The TLL model provides a theoretical framework to describe the low-energy excitations of interacting electrons in 1D systems. A hallmark of TLL behavior is the emergence of spin-charge separation, where the spin and charge degrees of freedom form distinct collective excitations, known as spinons and holons. To investigate TLL behavior in CuTe/Cu(111), we analyzed the density of states near the Fermi level, as shown in Fig. 4(e) and 4(f). The data reveal a power-law scaling up to 300 K that deviates from the standard Fermi-Dirac distribution, consistent with the predictions of the TLL model [see also Fig. S14]. From this analysis, we estimate the scaling factor α to be approximately 0.62, which is comparable to values in other TLL systems [43,44]. This quantitative result provides strong evidence of spin-charge separation in the CuTe chains.

The discovery of 1D flat bands and temperature-driven phase transition in the zigzag lattice offers new opportunities to explore intriguing physical properties that arise from the interplay between electron-correlation effects, lattice geometry, and quantum confinements. Although it is difficult to perform direct transport measurements on the CuTe/Cu(111) system, it might be suitable for (ultrafast) optical or plasmonic devices since electronic states that are far from the Fermi level plays a crucial role in determining the optical and plasmonic properties of materials. For example, the flat valence and conduction bands can enable triplet excitonic insulator [45] and stabilize the excitonic condensate [46].

In summary, we predicted and verified the existence of 1D flat bands in CuTe chains that host the 1D zigzag lattice. In addition, we observed a temperature-driven phase transition at approximately 250 K. Our detailed analyses suggest that the system has a TLL behavior, accompanied by spin-charge separation. Our results offer new opportunities to explore intriguing physical properties that arise from the interplay between electron-correlation effects, lattice geometry, and quantum confinements.

**Acknowledgments**
This work was supported by the National Key R&D Program of China (Grant No. 2024YFA1408400), the Beijing Natural Science Foundation (Grant No. JQ23001), the National Natural Science Foundation of China (Grants No. 12374197 and No. W2411004), the Strategic Priority Research Program of Chinese Academy of Sciences (Grants No. XDB33030100), and the CAS Project for Young Scientists in Basic Research (Grant No. YSBR-047). The synchrotron ARPES experiments were performed with the approval of the Proposal Assessing Committee of the Hiroshima Synchrotron Radiation Center (Proposals No. 23AG024 and No. 23AG025).

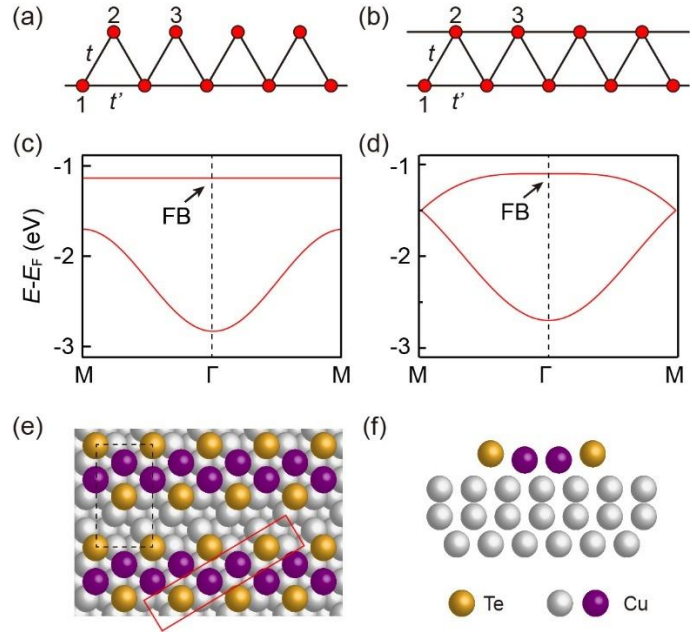

**Fig. 1:** (a,b) Schematic illustration of the 1D sawtooth and zigzag lattices, respectively. $t$ and $t'$ correspond to the hopping of an electron from one atom to its nearest neighbor along the two different bond directions. (c,d) Calculated band structure of the sawtooth ($t = 0.4\ eV$, $t' = 0.4/\sqrt{2}\ eV$, $U$=-1.7 eV) and zigzag ($t$=0.4 eV, $t'$=0.1 eV, $U$=-1.7 eV) lattices, respectively. The flat band is indicated by black arrows. (e) Top view of CuTe chains on Cu(111). The purple and orange balls represent Cu and Te atoms in CuTe chains, respectively, and the gray balls represent Cu atoms of the Cu(111) substrate. The black dashed square indicates the unit cell of ordered CuTe chains on Cu(111). (f) Side view of CuTe chains on Cu(111) from the red rectangle in (e).

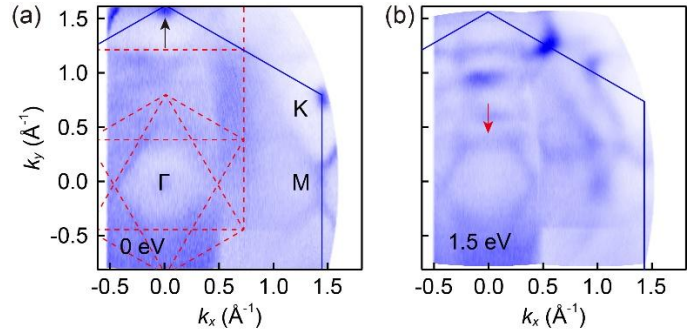

**Fig. 2:** CECs of CuTe/Cu(111) at binding energies of 0 eV and 1.5 eV, respectively. Blue lines indicate the first BZ of Cu(111); red dashed lines indicate the BZs of the $3 \times \sqrt{3}$ superstructure. The black arrow in (a) indicates a hole-like band centered at the Γ point of CuTe or K point of Cu(111). The Red arrow in (b) indicates the flat band of CuTe, which forms a hexagon because of the coexistence of three rotational domains.

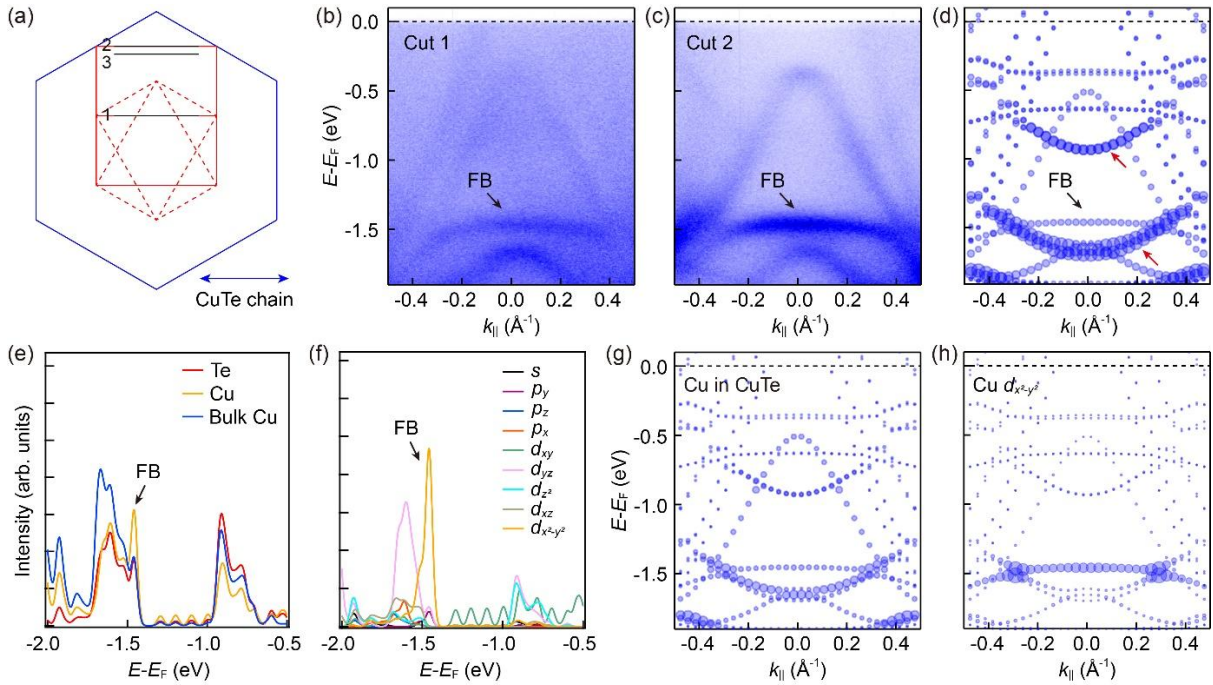

**Fig. 3:** (a) Schematic drawings of the BZs of Cu(111) (blue) and the $3 \times \sqrt{3}$ superstructure (red), respectively. The solid black lines indicate 3 momentum cuts discussed in the following. (b,c) ARPES intensity plots along Cuts 1 and 2, as indicated in (a). (d) Calculated band structures along Cut 1, unfolded to the first BZ of Cu(111). The marker size represents the spectral weight projected onto the CuTe and topmost Cu layer. (e) The spectral weight of different components (per atom) at various energies within the momentum range of the flat band. The original data was convoluted with a Gaussian function to simulate the broadening of the peaks. (f) The same as

(e) but for different orbitals of Cu in CuTe. (g,h) The same as (d), but the marker size represents the spectral weight projected onto the Cu atoms in CuTe and $d_{x^2-y^2}$ orbitals of Cu, respectively.

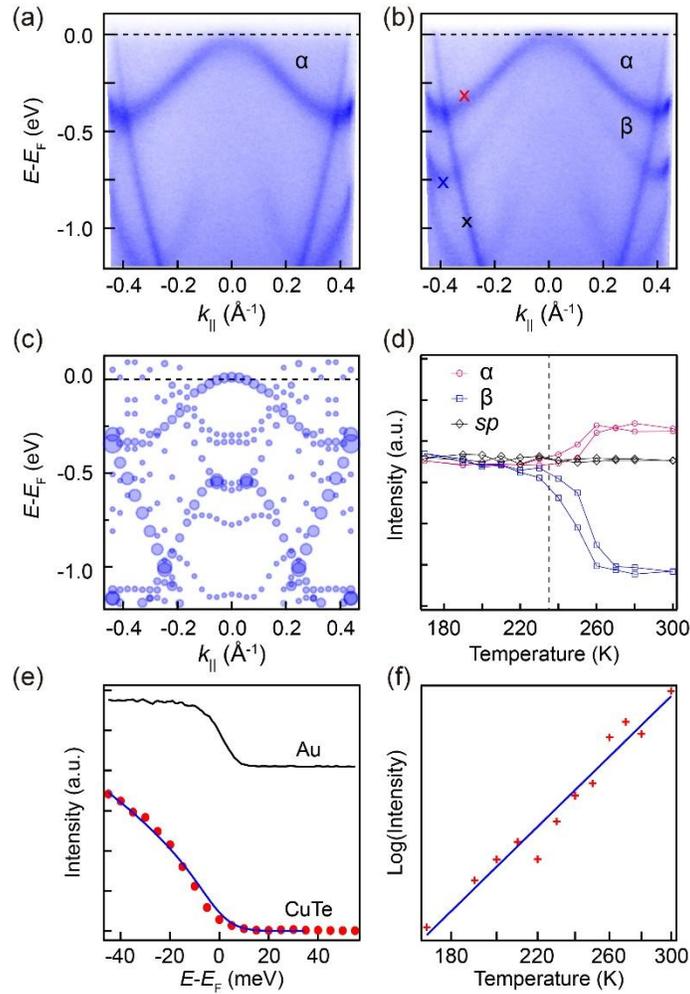

**Fig. 4:** (a) and (b) ARPES intensity plots along the $\overline{KMK}$ direction of Cu(111) measured at 300 K and 70 K, respectively. (c) Calculated band structures along the same cut. (d) Spectral weight of the α, β, and Cu sp bands, respectively, as a function of temperature. (e) Energy distribution curves near the Fermi level for polycrystalline Au and CuTe/Cu(111), respectively. The blue line is the fitting curve. (f) Temperature dependence of the photoelectron intensity near the Fermi level in a log scale. The blue line is the fitting curve based on power-law scaling (0.62).